\documentclass[runningheads,orivec]{llncs}

\def\cameraready{true}
\def\pdfmetadata{true}

\input{vars}

\usepackage[T1]{fontenc}
\usepackage[utf8]{inputenc}

\usepackage[
  english
]{babel}

\usepackage[
  pdftex
]{graphicx}
\usepackage{wrapfig}
\usepackage{adjustbox}
\usepackage{calc}

\usepackage{booktabs}
\usepackage{tabularx}
\usepackage{multirow}
\usepackage{multicol}

\usepackage{calc}

\usepackage[
  nolist
]{acronym}

\usepackage{color}
\usepackage[
  table
]{xcolor}
\usepackage[most]{tcolorbox}

\usepackage{tikz}
\usetikzlibrary{
  positioning,
  calc,
  arrows,
  arrows.meta,
  shapes.misc,
  shapes.geometric,
  decorations.markings,
  patterns,
  patterns.meta
}

\usepackage{tikz-layers}

\usepackage{listings}

\usepackage{xspace}

\usepackage{amsmath}
\usepackage{amssymb}
\usepackage{marvosym}

\usepackage{enumitem}

\usepackage{siunitx}

\def\papertitle{Relevant HAL Interface Requirements for~Embedded Systems}

\title{\papertitle{}}
\pdftitle{\papertitle{}}

\authorrunning{\censor{Bentele, Podelski, Sikora, Westphal}}
\author{\censor{Manuel~Bentele\inst{1,3}\orcidID{0009-0003-4794-958X}} \and
        \censor{Andreas~Podelski\inst{1}\orcidID{0000-0003-2540-9489}} \and \\
        \censor{Axel~Sikora\inst{2,3}\orcidID{0000-0003-0878-2919}} \and
        \censor{Bernd~Westphal\inst{4}\orcidID{0000-0002-6824-0567}}}
\pdfauthor{Manuel Bentele, Andreas Podelski, Axel Sikora, Bernd Westphal}
\institute{\censor{University of Freiburg, Freiburg, Germany} \and
           \censor{Offenburg University of Applied Sciences, Offenburg, Germany} \and
           \censor{Hahn-Schickard Institute, Villingen-Schwenningen, Germany} \and
           \censor{German Aerospace Center, Oldenburg, Germany}}

\input{modes}

\definecolor{islamicgreen}{rgb}{0.0, 0.56, 0.0}

\colorlet{cgreen}{green}
\colorlet{cdgreen}{islamicgreen}
\colorlet{cyellow}{yellow}
\colorlet{corange}{orange}
\colorlet{cred}{red}
\colorlet{cviolet}{violet}
\colorlet{cblue}{cyan}
\colorlet{cdlue}{blue}
\colorlet{cgray}{gray}
\colorlet{cblack}{black}

\colorlet{bgray}{gray!10}
\colorlet{lgray}{gray!5}

\newcommand{\bFillOpacity}{0.15}
\newcommand{\bFillIntensity}{15}

\newcommand*{\belowrulesepcolor}[1]{%
    \noalign{%
        \kern-\belowrulesep
        \begingroup
        \color{#1}%
        \hrule height\belowrulesep
        \endgroup
    }%
}

\newcommand*{\aboverulesepcolor}[1]{%
    \noalign{%
        \begingroup
        \color{#1}%
        \hrule height\aboverulesep
        \endgroup
        \kern-\aboverulesep
    }%
}

\tikzset{
    box/.style = {draw, rectangle, fill = none},
    font = \footnotesize,
    thadRoutine/.style={minimum width=10pt, minimum height=10pt},
    thadRoutineBox/.style={thadRoutine,draw, rectangle, fill=white,inner sep=3pt,text centered},
    thad/.style={preaction={on layer=background,line width=10pt,draw=cblue!\bFillIntensity},
        draw=cgray,
        decoration = { markings, mark=at position #1 with {
                \node[anchor=center,transform shape,inner sep=2.5pt, fill=cblue!\bFillIntensity] {$\triangleleft$}; }},
        postaction={decorate} },
    thad/.default=0.5,
    arrow/.style = {-triangle 45},
    crossNode/.style = {draw, fill = black, solid, circle, minimum size = 1mm, inner sep = 0pt},
    device/.style = {box, minimum width = 2.5cm, minimum height = 1.5cm, inner sep = 0pt},
    slave/.style = {fill = cgray, fill opacity = 0.15},
    pics/slaveFourInputsStyle/.style args = {#1/#2/#3/#4/#5}{
        code = {
            \node[device, slave, #5] (-sl) {};
            \node[anchor = east] at (-sl.east) {\tikzpictext};
            \coordinate (-p1) at ($(-sl.north west)!0.2!(-sl.south west)$);
            \coordinate (-p2) at ($(-sl.north west)!0.4!(-sl.south west)$);
            \coordinate (-p3) at ($(-sl.north west)!0.6!(-sl.south west)$);
            \coordinate (-p4) at ($(-sl.north west)!0.8!(-sl.south west)$);
            \node[font=\tiny,anchor=west] (-1) at (-p1) {#1};
            \node[font=\tiny,anchor=west] (-2) at (-p2) {#2};
            \node[font=\tiny,anchor=west] (-3) at (-p3) {#3};
            \node[font=\tiny,anchor=west] (-4) at (-p4) {#4};
        }
    },
    pics/masterFourInputs/.style args = {#1/#2/#3/#4}{
        code = {
            \node[device, master] (-ma) {};
            \node[anchor = west] at (-ma.west) {\tikzpictext};
            \coordinate (-p1) at ($(-ma.north east)!0.2!(-ma.south east)$);
            \coordinate (-p2) at ($(-ma.north east)!0.4!(-ma.south east)$);
            \coordinate (-p3) at ($(-ma.north east)!0.6!(-ma.south east)$);
            \coordinate (-p4) at ($(-ma.north east)!0.8!(-ma.south east)$);
            \node[font=\tiny,anchor=east] (-1) at (-p1) {#1};
            \node[font=\tiny,anchor=east] (-2) at (-p2) {#2};
            \node[font=\tiny,anchor=east] (-3) at (-p3) {#3};
            \node[font=\tiny,anchor=east] (-4) at (-p4) {#4};
        }
    },
    pics/masterSevenInputs/.style args = {#1/#2/#3/#4/#5/#6/#7}{
        code = {
            \node[device, master, minimum height = 2.4cm] (-ma) {};
            \node[anchor = west] at (-ma.west) {\tikzpictext};
            \coordinate (-p1) at ($(-ma.north east)!0.125!(-ma.south east)$);
            \coordinate (-p2) at ($(-ma.north east)!0.250!(-ma.south east)$);
            \coordinate (-p3) at ($(-ma.north east)!0.375!(-ma.south east)$);
            \coordinate (-p4) at ($(-ma.north east)!0.500!(-ma.south east)$);
            \coordinate (-p5) at ($(-ma.north east)!0.625!(-ma.south east)$);
            \coordinate (-p6) at ($(-ma.north east)!0.750!(-ma.south east)$);
            \coordinate (-p7) at ($(-ma.north east)!0.875!(-ma.south east)$);
            \node[font=\tiny,anchor=east] (-1) at (-p1) {#1};
            \node[font=\tiny,anchor=east] (-2) at (-p2) {#2};
            \node[font=\tiny,anchor=east] (-3) at (-p3) {#3};
            \node[font=\tiny,anchor=east] (-4) at (-p4) {#4};
            \node[font=\tiny,anchor=east] (-5) at (-p5) {#5};
            \node[font=\tiny,anchor=east] (-6) at (-p6) {#6};
            \node[font=\tiny,anchor=east] (-7) at (-p7) {#7};
        }
    },
    pics/swlapp/.style args = {#1}{
        code = {
            \node[minimum width=0.4\textwidth, minimum height=0.03\textheight] (-app) {#1};
            \coordinate (-c11) at ($(-app.north west) + (0, 0)$);
            \coordinate (-c12) at ($(-app.south west) + (0, -0.02\textheight)$);
            \coordinate (-c13) at ($(-app.south west) + (0.05\textwidth-3pt, -0.02\textheight)$);
            \coordinate (-c14) at ($(-app.south west) + (0.05\textwidth-3pt, 0)$);
            \coordinate (-c21) at ($(-app.north east) + (0, 0)$);
            \coordinate (-c22) at ($(-app.south east) + (0, -0.02\textheight)$);
            \coordinate (-c23) at ($(-app.south east) + (-0.05\textwidth+3pt, -0.02\textheight)$);
            \coordinate (-c24) at ($(-app.south east) + (-0.05\textwidth+3pt, 0)$);
            \begin{pgfonlayer}{behind}
                \draw[fill = cred,fill opacity=\bFillOpacity] (-c11) -- (-c12) -- (-c13) -- (-c14) -- (-c24) -- (-c23) -- (-c22) -- (-c21) -- cycle;
            \end{pgfonlayer}
        }
    },
    pics/swlapp/.default=Application,
    pics/swlint/.style args = {#1/#2/#3}{
        code = {
            \node[minimum width=0.4\textwidth, minimum height=0.08\textheight] (-oshal) {};
            \coordinate (-c10) at ($(-oshal.south west) + (0,0)$);
            \coordinate (-c11) at ($(-oshal.south west) + (0,0.03\textheight)$);
            \coordinate (-c12) at ($(-oshal.north west) + (0,-0.02\textheight)$);
            \coordinate (-c13) at ($(-oshal.north west) + (0.05\textwidth,-0.02\textheight)$);
            \coordinate (-c14) at ($(-oshal.north west) + (0.05\textwidth,0)$);
            \coordinate (-c20) at ($(-oshal.south east) + (0,0)$);
            \coordinate (-c21) at ($(-oshal.south east) + (0,0.03\textheight)$);
            \coordinate (-c22) at ($(-oshal.north east) + (0,-0.02\textheight)$);
            \coordinate (-c23) at ($(-oshal.north east) + (-0.05\textwidth,-0.02\textheight)$);
            \coordinate (-c24) at ($(-oshal.north east) + (-0.05\textwidth,0)$);
            \draw (-c13) -- (-c12) -- (-c11) -- (-c10) -- (-c20) -- (-c21) -- (-c22) -- (-c23);
            \draw[dashed] (-c11) -- (-c21);
            \draw[dashed] (-c13) -- (-c14) -- (-c24) -- (-c23) -- cycle;
            \begin{pgfonlayer}{behind}
                \fill[fill = cblue, fill opacity=\bFillOpacity] (-c11) -- (-c12) -- (-c22) -- (-c21);
                \fill[fill = cgreen, fill opacity=\bFillOpacity] (-c13) -- (-c14) -- (-c24) -- (-c23);
                \fill[fill = corange, fill opacity=\bFillOpacity] (-c10) -- (-c11) -- (-c21) -- (-c20);
            \end{pgfonlayer}
            \node[anchor=center] at ($(-oshal.south) + (0,0.07\textheight)$) {#1};
            \node[anchor=center] at ($(-oshal.south) + (0,0.045\textheight)$) {#2};
            \node[anchor=center] at ($(-oshal.south) + (0,0.015\textheight)$) {#3};
        }
    },
    pics/swlint/.default=API/Implementation/OS~SPI~subsystem,
    pics/swlintst/.style args = {#1/#2}{
        code = {
            \node[minimum width=0.4\textwidth, minimum height=0.05\textheight] (-midhal) {};
            \coordinate (-c100) at ($(-midhal.south west) + (0.05\textwidth-3pt,0)$);
            \coordinate (-c10) at ($(-midhal.south west) + (0.05\textwidth-3pt,-0.02\textheight)$);
            \coordinate (-c11) at ($(-midhal.south west) + (0,-0.02\textheight)$);
            \coordinate (-c12) at ($(-midhal.north west) + (0,-0.02\textheight)$);
            \coordinate (-c13) at ($(-midhal.north west) + (0.05\textwidth,-0.02\textheight)$);
            \coordinate (-c14) at ($(-midhal.north west) + (0.05\textwidth,0)$);
            \coordinate (-c200) at ($(-midhal.south east) + (-0.05\textwidth+3pt,0)$);
            \coordinate (-c20) at ($(-midhal.south east) + (-0.05\textwidth+3pt,-0.02\textheight)$);
            \coordinate (-c21) at ($(-midhal.south east) + (0,-0.02\textheight)$);
            \coordinate (-c22) at ($(-midhal.north east) + (0,-0.02\textheight)$);
            \coordinate (-c23) at ($(-midhal.north east) + (-0.05\textwidth,-0.02\textheight)$);
            \coordinate (-c24) at ($(-midhal.north east) + (-0.05\textwidth,0)$);
            \draw (-c13) -- (-c12) -- (-c11) -- (-c10) -- (-c100) -- (-c200) -- (-c20) -- (-c21) -- (-c22) -- (-c23);
            \draw[dashed] (-c13) -- (-c14) -- (-c24) -- (-c23) -- cycle;
            \begin{pgfonlayer}{behind}
                \fill[fill = cblue, fill opacity=\bFillOpacity] (-c100) -- (-c10) -- (-c11) -- (-c12) -- (-c22) -- (-c21) -- (-c20) -- (-c200);
                \fill[fill = cgreen, fill opacity=\bFillOpacity] (-c13) -- (-c14) -- (-c24) -- (-c23);
            \end{pgfonlayer}
            \node[anchor=center] at ($(-midhal.south) + (0,0.04\textheight)$) {#1};
            \node[anchor=center] at ($(-midhal.south) + (0,0.015\textheight)$) {#2};
        }
    },
    pics/swlintst/.default=API/Implementation,
    pics/swlintend/.style args = {#1/#2}{
        code = {
            \node[minimum width=0.4\textwidth, minimum height=0.05\textheight] (-endhal) {};
            \coordinate (-c11) at ($(-endhal.south west) + (0,0)$);
            \coordinate (-c12) at ($(-endhal.north west) + (0,-0.02\textheight)$);
            \coordinate (-c13) at ($(-endhal.north west) + (0.05\textwidth,-0.02\textheight)$);
            \coordinate (-c14) at ($(-endhal.north west) + (0.05\textwidth,0)$);
            \coordinate (-c21) at ($(-endhal.south east) + (0,0)$);
            \coordinate (-c22) at ($(-endhal.north east) + (0,-0.02\textheight)$);
            \coordinate (-c23) at ($(-endhal.north east) + (-0.05\textwidth,-0.02\textheight)$);
            \coordinate (-c24) at ($(-endhal.north east) + (-0.05\textwidth,0)$);
            \draw (-c11) -- (-c12) -- (-c13) -- (-c14) -- (-c24) -- (-c23) -- (-c22) -- (-c21) -- cycle;
            \draw[dashed] (-c13) -- (-c23);
            \begin{pgfonlayer}{behind}
                \fill[fill = cblue, fill opacity=\bFillOpacity] (-c11) -- (-c12) -- (-c22) -- (-c21);
                \fill[fill = cgreen, fill opacity=\bFillOpacity] (-c13) -- (-c14) -- (-c24) -- (-c23);
            \end{pgfonlayer}
            \node[anchor=center] at ($(-endhal.south) + (0,0.04\textheight)$) {#1};
            \node[anchor=center] at ($(-endhal.south) + (0,0.015\textheight)$) {#2};
        }
    },
    pics/swlintst/.default=API/Implementation,
    pics/swlhw/.style args = {#1}{
        code = {
            \node[draw,fill=cgray,fill opacity=\bFillOpacity, text opacity=1, minimum width=0.4\textwidth, minimum height = 0.03\textheight] (-hw) {#1};
        }
    },
    pics/swlhw/.default=Hardware
}

\crefname{cntex}{Example}{Examples}
\newtcolorbox[use counter=cntex]{examplerequirement}[2]{
  title={\textbf{Example~\thetcbcounter}: #1},
  label={#2},
  standard jigsaw,
  opacityback=0,
  sharp corners,
  colbacktitle=gray,
  boxrule=0.5pt,
  boxsep=3pt,
  left=5pt,
  right=5pt
}

\crefname{cntes}{Embedded system}{Embedded systems}
\newtcolorbox[use counter=cntes]{embeddedsystem}[2]{
  title={\textbf{Embedded~system~\thetcbcounter}: #1},
  label={#2},
  breakable,
  enhanced,
  enhanced jigsaw,
  opacityback=0,
  sharp corners,
  colbacktitle=gray,
  boxrule=0.5pt,
  boxsep=3pt,
  left=5pt,
  right=5pt
}

\newlength{\enummarginlr}
\newlength{\enumsep}
\newlist{researchquestions}{enumerate}{1}
\setlist[researchquestions]{%
  label={\textbf{RQ\arabic*}:},
  leftmargin=*,
  labelindent=\enummarginlr,
  rightmargin=\enummarginlr,
  itemsep=\enumsep
}

\usepackage{url}

\gappto{\UrlBreaks}{\UrlOrds}

\ifdef{\pdfmetadata}{
  
  \urlstyle{rm}
}{}

\ifdef{\cameraready}{
  \newcommand{\censor}[1]{#1}
}{
  \usepackage{censor}
}

\makeatletter

\DeclareRobustCommand\onedot{\futurelet\@let@token\@onedot}
\newcommand*{\@onedot}{\ifx\@let@token.\else.\null\fi\xspace}

\newcommand*{\textoverline}[1]{$\overline{\hbox{#1}}\m@th$}

\def\lst@makecaption{%
    \def\@captype{table}%
    \@makecaption
}

\makeatother

\BeforeBeginEnvironment{wrapfigure}{\setlength{\intextsep}{1em}}

\newcommand{\eg}{e.g\onedot}

\newcommand{\ie}{i.e\onedot}

\newcommand{\cf}{cf\onedot}

\newcommand{\singlequote}[1]{`#1'}
\newcommand{\doublequote}[1]{``#1''}

\newcommand{\numunit}[2]{#1\,#2}

\newcommand{\code}[1]{\texttt{#1}\xspace}
\newcommand{\codeParam}[1]{\code{#1}}

\newcommand{\halApiRoutineOpen}{\code{open()}}
\newcommand{\halApiRoutineOpenS}{\code{open}}

\newcommand{\halApiRoutineClose}{\code{close()}}

\newcommand{\halApiRoutineRead}{\code{read()}}
\newcommand{\halApiRoutineReadS}{\code{read}}

\newcommand{\halApiRoutineWrite}{\code{write()}}
\newcommand{\halApiRoutineWriteS}{\code{write}}

\newcommand{\halApiRoutineIoctl}[1]{\code{ioctl(#1)}}
\newcommand{\halApiRoutineIoctlS}{\code{ioctl}}

\newcommand{\halApiRoutineParamValueMsg}{\codeParam{MSG}}

\newcommand{\halApiRoutineParamValueRMode}{\codeParam{RD\_MODE}}
\newcommand{\halApiRoutineParamValueWMode}{\codeParam{WR\_MODE}}

\newcommand{\halApiRoutineParamValueRModeExt}{\codeParam{RD\_MODE32}}
\newcommand{\halApiRoutineParamValueWModeExt}{\codeParam{WR\_MODE32}}

\newcommand{\halApiRoutineParamValueRLSBF}{\codeParam{RD\_LSB\_FIRST}}
\newcommand{\halApiRoutineParamValueWLSBF}{\codeParam{WR\_LSB\_FIRST}}

\newcommand{\halApiRoutineParamValueRBitsPWord}{\codeParam{RD\_BITS\_PER\_WORD}}
\newcommand{\halApiRoutineParamValueWBitsPWord}{\codeParam{WR\_BITS\_PER\_WORD}}

\newcommand{\halApiRoutineParamValueRSpeed}{\codeParam{RD\_MAX\_SPEED\_HZ}}
\newcommand{\halApiRoutineParamValueWSpeed}{\codeParam{WR\_MAX\_SPEED\_HZ}}

\newcommand{\halApiRoutineIoctlMsg}{\halApiRoutineIoctl{\halApiRoutineParamValueMsg}}

\newcommand{\halApiRoutineIoctlRMode}{\halApiRoutineIoctl{\halApiRoutineParamValueRMode}}

\newcommand{\halApiRoutineIoctlWMode}{\halApiRoutineIoctl{\halApiRoutineParamValueWMode}}

\newcommand{\halApiRoutineIoctlRModeExt}{\halApiRoutineIoctl{\halApiRoutineParamValueRModeExt}}

\newcommand{\halApiRoutineIoctlWModeExt}{\halApiRoutineIoctl{\halApiRoutineParamValueWModeExt}}

\newcommand{\halApiRoutineIoctlRLSBF}{\halApiRoutineIoctl{\halApiRoutineParamValueRLSBF}}

\newcommand{\halApiRoutineIoctlWLSBF}{\halApiRoutineIoctl{\halApiRoutineParamValueWLSBF}}

\newcommand{\halApiRoutineIoctlRBitsPWord}{\halApiRoutineIoctl{\halApiRoutineParamValueRBitsPWord}}

\newcommand{\halApiRoutineIoctlWBitsPWord}{\halApiRoutineIoctl{\halApiRoutineParamValueWBitsPWord}}

\newcommand{\halApiRoutineIoctlRSpeed}{\halApiRoutineIoctl{\halApiRoutineParamValueRSpeed}}

\newcommand{\halApiRoutineIoctlWSpeed}{\halApiRoutineIoctl{\halApiRoutineParamValueWSpeed}}

\newcommand{\opdep}{\triangleleft}

\newcommand{\varThad}{\delta}
\newcommand{\varThadDep}[3]{#1 : #2 \opdep #3}

\newcommand{\cbrackets}[1]{\left\{#1 \right\}}
\newcommand{\set}[1]{\cbrackets{#1}}

\newcommand{\spidev}{\emph{spidev}\xspace}
\newcommand{\ultimateAutomizer}{\textsc{Ultimate Automizer}\xspace}

\ifdef{\draftmode}{
  \usepackage{todonotes}
}{
  \usepackage[
    disable
  ]{todonotes}
}

\makeatletter
\AtBeginDocument
 {
   \def\ltx@label#1{\cref@label{#1}}
   \def\label@in@display@noarg#1{\cref@old@label@in@display{#1}}
   \def\label@in@mmeasure@noarg#1{%
    \begingroup%
      \measuring@false%
      \cref@old@label@in@display{#1}
    \endgroup}%
 } %
\makeatother

\begin{document}

  \maketitle

\begin{acronym}
  \acro{api}[API]{Application Programming Interface}
  \acroplural{api}[APIs]{Application Programming Interfaces}
  \acro{hal}[HAL]{Hardware Abstraction Layer}
  \acroplural{hal}[HALs]{Hardware Abstraction Layers}
  \acro{thad}[THAD]{Temporal HAL-API Dependency}
  \acroplural{thad}[THADs]{Temporal HAL-API Dependencies}
  \acro{ast}[AST]{Abstract Syntax Tree}
  \acroplural{ast}[ASTs]{Abstract Syntax Trees}
  \acro{ltl}[LTL]{Linear Temporal Logic}
  \acro{acsl}[ACSL]{ANSI/ISO C~Specification Language}
  \acro{pin}[PIN]{Personal Identification Number}
  \acroplural{pin}[PINs]{Personal Identification Numbers}
  \acro{usb}[USB]{Universal Serial Bus}
  \acroplural{usb}[USBs]{Universal Serial Busses}
  \acro{spi}[SPI]{Serial Peripheral Interface}
  \acroplural{spi}[SPIs]{Serial Peripheral Interfaces}
  \acro{cpu}[CPU]{Central Processing Unit}
  \acroplural{cpu}[CPUs]{Central Processing Units}
  \acro{sck}[SCK]{Serial Clock}
  \acro{mosi}[MOSI]{Master Out/Slave In}
  \acro{miso}[MISO]{Master In/Slave Out}
  \acro{ss}[SS]{Slave Select}
  \acro{cpol}[CPOL]{Clock Polarity}
  \acro{cpha}[CPHA]{Clock Phase}
  \acro{lsb}[LSB]{Least Significant Bit}
  \acro{msb}[MSB]{Most Significant Bit}
  \acro{lsbfe}[LSBFE]{Least Significant Bit First Enable}
  \acro{vfs}[VFS]{Virtual File System}
  \acro{posix}[POSIX]{Portable Operating System Interface}
  \acro{svcomp}[SV-COMP]{Competition on Software Verification}
  \acro{slam}[SLAM]{Software (Specifications), programming Languages, Abstraction, and Model checking}
  \acro{sdv}[SDV]{Static Driver Verifier}
  \acro{slic}[SLIC]{Specification Language for Interface Checking}
  \acro{misra}[MISRA]{Motor Industry Software Reliability Association}
  \acro{smt}[SMT]{Satisfiability Modulo Theories}
  \acro{loc}[LOC]{Lines of Code}
  \acro{hw}[HW]{Hardware}
  \acro{sw}[SW]{Software}
  \acro{id}[ID]{Identity}
  \acused{api}
  \acused{pin}
  \acused{usb}
  \acused{spi}
  \acused{cpu}
  \acused{sck}
  \acused{miso}
  \acused{mosi}
  \acused{ss}
  \acused{cpol}
  \acused{cpha}
  \acused{lsb}
  \acused{msb}
  \acused{lsbfe}
  \acused{posix}
  \acused{svcomp}
  \acused{slam}
  \acused{slic}
  \acused{misra}
  \acused{smt}
  \acused{loc}
  \acused{hw}
  \acused{sw}
  \acused{id}
\end{acronym}


\newcommand{\abstractparagraph}[1]{\textbf{[#1]}}

\begin{abstract}
\abstractparagraph{Context \& Motivation}
Embedded applications often use a \ac{hal} to access hardware.
Improper use of the \ac{hal} can lead to incorrect hardware operations, resulting in system failure and potentially serious damage to the hardware.
\abstractparagraph{Question \& Problem}
The question is how one can obtain prioritize, among a possibly large set of \ac{hal} interface requirements, those that are indisputably relevant for preventing this kind of system failure.
\abstractparagraph{Ideas \& Results}
In this paper, we introduce a formal notion of relevance.
This allows us to leverage a formal method, \ie, software model checking, to produce a mathematical proof that a requirement is indisputably relevant.
We propose an approach to extract provably relevant requirements from issue reports on system failures.
We present a case study to demonstrate that the approach is feasible in principle.
The case study uses three examples of issue reports on embedded applications that use the \ac{spi} bus via the \spidev \ac{hal}.
\abstractparagraph{Contribution}
The overall contribution of this paper is to pave the way for the study of approaches to a new kind of prioritization aimed at preventing a specific kind of system failure.
\pdfkeywords{Requirements Prioritization, Embedded Systems, Hardware Abstraction Layer, HAL Interface Requirements, Serial Peripheral Interface (SPI), Formal Methods, Software Model Checking}
\end{abstract}


\section{Introduction}
\label{sec:introduction}
Developing application code that uses a \ac{hal} to access hardware of an embedded system poses specific requirements engineering challenges.
Besides the usual system and software requirements, which are typically defined at the start of a project, \ac{hal} interface requirements are often ignored until later, perhaps too late:
The improper use of the \ac{hal} can lead to incorrect hardware operations resulting in system failure and potentially serious hardware damage.

Unlike system requirements, which vary depending on the specific application, interface requirements are determined by the \ac{hal} and can be reused across multiple system designs.
There are, however, many of those.
The fact that the number of \ac{hal} interface requirements can be large and overwhelming, may be one reason why they are often ignored until their violation has indeed caused a system failure.
This raises naturally the question how one can prioritize, among a possibly large set of \ac{hal} interface requirements, those that are indisputably relevant for preventing this kind of system failure that can lead to potentially serious hardware damage.

Requirements prioritization plays an important role in requirements engineering~\cite{Hujainah2018,Bukhsh2020,Yaseen2025}.
Prioritization requires a criterion that can be used to select a requirement.
The problem is that a criterion often lacks a clear understanding, which makes prioritization subjective and inconsistent~\cite{Hujainah2018,Bukhsh2020,Yaseen2025}.

In this paper, we address exactly this problem and introduce a novel criterion which can be defined  precisely and unambiguously.
The criterion of relevance can be used to prioritize requirements that are {indisputably} relevant for preventing system failures.
We call this criterion \emph{relevant}, for short.
The definition of relevance allows us to leverage a formal method, \ie, software model checking~\cite{Heizmann2013},  to produce a mathematical proof that a requirement is relevant (and, in this sense, indisputably relevant).

We propose an approach to extract provably relevant requirements from issue reports on system failures.
The approach comes in two versions.
In the first version, one infers a candidate requirement from the issue report and then applies software model checking to check whether the candidate requirement is a relevant requirement, or not.
In the second version, one assumes that one is given a (possibly large) set of \ac{hal} interface requirements for, say, a specific \ac{hal}.
The approach is then to take the whole set as a set of candidate requirements and apply software model checking to each of them.  In the second version, the approach relies on the fact that the application of software model checking as a formal method is fully automatic.

We present a preliminary case study to demonstrate that the approach is feasible in principle.
The case study uses three examples of issue reports on embedded applications that use the \ac{spi} bus via the \spidev \ac{hal}.
For the second version of the approach, the case study uses a set of \num{26} \ac{hal} interface requirements for the \spidev \ac{hal} (and checks each of those on the two versions of each of the three application programs).
The requirements had to be extracted manually.
They form the set of all \ac{hal} interface requirements for the \spidev \ac{hal} that are of a specific form.

The contribution of this paper is to define a precise and unambiguous criterion that can be used for prioritization, to propose an approach to use that criterion, and to present a case study to demonstrate that the approach is feasible in principle.
The overall contribution is perhaps to pave the way for the study of approaches to a new kind of prioritization aimed at preventing a specific kind of system failure.

The paper is structured as follows:
In \Cref{sec:approach-process}, we introduce our approach.
In \Cref{sec:relevant-requirements}, we present the experiments to evaluate the approach.
In \Cref{sec:related-work}, we discuss related work.
\Cref{sec:conclusion} is the conclusion section.


\section{An approach based on relevance}
\label{sec:approach-process}
In this section, we introduce a formal notion of relevance and an approach.
Given an issue report on a system failure, we present two variants of the approach: one that extracts a single interface requirement from the issue and checks its relevance, and the other checks a given set of candidate requirements for relevance.

\subsection{Notion of relevance}
Intuitively, we consider a requirement to be relevant if its validity is essential to prevent a system failure.
\Cref{def:relevant-requirement} captures this intuition precisely.

\begin{definition}[Relevance]\label{def:relevant-requirement}
A requirement~$R$ is \emph{relevant} if there exists a system that comes in two versions: a faulty version, \ie, a system that exhibits a system failure, and a repaired version, \ie, a system that no longer exhibits the system failure, such that the original application program in the faulty system violates~$R$ and the new application program in the repaired system satisfies~$R$.
\end{definition}

\let\oldintextsep\intextsep%
\setlength\intextsep{-\intextsep}%
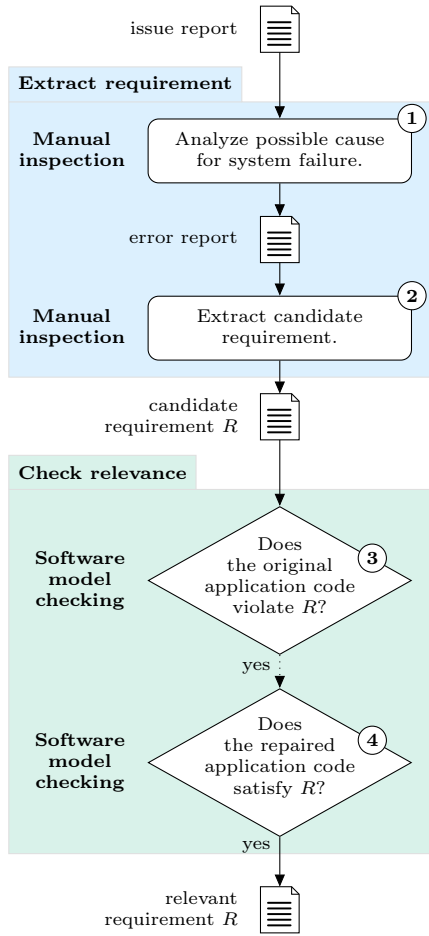
\begin{wrapfigure}[34]{L}{6cm}
  \centering

\newlength{\processnodedistance}
\setlength{\processnodedistance}{0.44cm}
\newlength{\processitemwidth}
\setlength{\processitemwidth}{3.5cm}
\newlength{\documentwidth}
\setlength{\documentwidth}{0.5cm}
\newlength{\documentheight}
\setlength{\documentheight}{0.6cm}
\newlength{\documentcorner}
\setlength{\documentcorner}{0.1cm}
\newcommand{\documentlines}{5}
\newlength{\documentlinewidth}
\setlength{\documentlinewidth}{0.075em}
\newlength{\documentmargin}
\setlength{\documentmargin}{0.1cm}
\newlength{\documenttopmargin}
\setlength{\documenttopmargin}{0.2cm}

\definecolor{mayablue}{rgb}{0.45, 0.76, 0.98}
\definecolor{mediumaquamarine}{rgb}{0.4, 0.8, 0.67}

\tikzset{%
  activity/.style={draw,rectangle,rounded corners,minimum width=\processitemwidth,text width=0.9\processitemwidth,align=flush center,font=\scriptsize,fill=white,inner sep=0.5em},
  decision/.style={draw,diamond,minimum width=\processitemwidth,text width=0.9\processitemwidth,aspect=1.6,font=\scriptsize,fill=white,inner sep=0em,text centered,align=center},
  number/.style={draw,circle,fill=white,inner sep=0.15em},
  label/.style={font=\scriptsize,align=right},
  heading/.style={font=\bfseries\scriptsize},
  info/.style={font=\tiny},
  arrow/.style={-{Triangle[angle=45:0.5em]}},
  pics/processactivity/.style n args={3}{
    code={
      \node[activity] (-box) {#3};
      \node[label,heading,anchor=east,xshift=-0.5em,align=center] (-label) at (-box.west) {#2};
      \node[number,heading,anchor=center] (-number) at (-box.north east) {#1};
    }
  },
  pics/processdecision/.style n args={3}{
    code={
      \node[decision] (-box) {#3};
      \node[label,heading,anchor=east,xshift=-0.5em,align=center] (-label) at (-box.west) {#2};
      \node[number,heading,anchor=center] (-number) at ($(-box.north east)!0.4!(-box.east)$) {#1};
    }
  },
  pics/processdocument/.style n args={1}{
    code={
      \node[minimum width=\documentwidth,minimum height=\documentheight] (-box) {};

      \coordinate (-sw) at (-box.south west);
      \coordinate (-nw) at (-box.north west);
      \coordinate (-ne) at (-box.north east);
      \coordinate (-se) at (-box.south east);
      \coordinate (-ne0) at ($(-ne) - (\documentcorner,0)$);
      \coordinate (-ne1) at ($(-ne) - (0,\documentcorner)$);

      \filldraw[fill=white] (-sw) -- (-nw) -- (-ne0) -- (-ne1) -- (-se) -- cycle;
      \draw (-ne0) -- (-ne0|--ne1) -- (-ne1);

      \foreach \k in {1,...,\documentlines}{
        \draw[line width=\documentlinewidth,line cap=round]
        ($(-sw) + (\documentmargin,\documentmargin) + (0,{(\k-1)/(\documentlines-1)*(\documentheight - \documentmargin - \documenttopmargin)})$) -- ++ ($(\documentwidth,0) - 2*(\documentmargin,0)$);
      }

      \node[label,anchor=east,xshift=-0.5em] at ($(-nw)!0.5!(-sw)$) {#1};
    }
  },
  node distance=\processnodedistance
}
\begin{tikzpicture}
  \pic (r0)                   {processdocument={issue report}};
  \pic [below=2\processnodedistance of r0-box] (a1) {processactivity={1}{Manual\\inspection}{Analyze possible cause for system failure.}};
  \pic (r1) [below=of a1-box] {processdocument={error report}};
  \pic [below=of r1-box] (a2) {processactivity={2}{Manual\\inspection}{Extract candidate requirement.}};
  \pic (r2) [below=of a2-box] {processdocument={candidate\\requirement~$R$}};
  \pic [below=2\processnodedistance of r2-box] (d1) {processdecision={3}{Software\\model\\checking}{}};
  \node[anchor=center,align=center,font=\scriptsize,yshift=0.2em] at (d1-box.center) {Does\\the original\\application code\\violate~$R$?};
  \pic [below=of d1-box] (d2) {processdecision={4}{Software\\model\\checking}{}};
  \node[anchor=center,align=center,font=\scriptsize,yshift=0.2em] at (d2-box.center) {Does\\the repaired\\application code\\satisfy~$R$?};
  \pic (r3) [below=1.5\processnodedistance of d2-box] {processdocument={relevant\\requirement~$R$}};

  \draw[arrow]        (r0-box) -- (a1-box);
  \draw[arrow]        (a1-box) -- (r1-box);
  \draw[arrow]        (r1-box) -- (a2-box);
  \draw[arrow]        (a2-box) -- (r2-box);
  \draw[arrow]        (r2-box) -- (d1-box);
  \draw[arrow,dotted] (d1-box) -- node[label,left,pos=0.4] {yes} (d2-box);
  \draw[arrow]        (d2-box) -- node[label,left,pos=0.2] {yes} (r3-box);

  \coordinate (anw) at ($(a1-box.north-|d1-label.west) + (-0.5\processnodedistance,0.5\processnodedistance)$);
  \coordinate (ase) at ($(a2-box.south east-|a2-number.east) + (0.25\processnodedistance,-0.5\processnodedistance)$);
  \node[heading,minimum height=\processnodedistance,anchor=south west,align=left] (g1) at (anw) {Extract requirement};
  \begin{pgfonlayer}{background}
    \draw[color=gray!25,fill=mayablue!25] (g1.north west) -- (g1.north east) -- (g1.south east) -- (g1.south east -| ase) -- (ase) -- (ase -| g1.north west) -- cycle;
    \draw[color=gray!25] (anw) -- (g1.south east);
  \end{pgfonlayer}

  \coordinate (dnw) at ($(d1-box.north-|d1-label.west) + (-0.5\processnodedistance,0.5\processnodedistance)$);
  \coordinate (dse) at ($(d2-box.south-|a2-number.east) + (0.25\processnodedistance,-0.5\processnodedistance)$);
  \node[heading,minimum height=\processnodedistance,anchor=south west,align=left] (g2) at (dnw) {Check relevance};
  \begin{pgfonlayer}{background}
    \draw[color=gray!25,fill=mediumaquamarine!25] (g2.north west) -- (g2.north east) -- (g2.south east) -- (g2.south east -| dse) -- (dse) -- (dse -| g2.north west) -- cycle;
    \draw[color=gray!25] (dnw) -- (g2.south east);
  \end{pgfonlayer}
\end{tikzpicture}
  \caption{Overall approach.
           The application code comes in two versions: the original version exhibits the system failure, the repaired version does not.}
  \label{fig:approach-process}
\end{wrapfigure}%
\setlength\intextsep{\oldintextsep}

For \Cref{def:relevant-requirement}, we note two remarks:
(1) The definition is based on an existential condition.
That is, we have a criterion that allows us to confirm relevance.
It does not allow us to conclude that an interface requirement will never become relevant.
(2) To make the definition practically useful, we need to apply it to existing system failures.
In theory, it may always be possible to build an artificial system that witnesses the relevance of a given interface requirement.

\subsection{Extract and check a single candidate requirement for relevance}
\label{sec:approach-process-overall}
The approach involves several stakeholders.
Hardware manufacturers are only partly involved.
They provide the \ac{hal}, particularly its documentation and occasionally the source code.
However, the extent and quality of the \ac{hal} documentation are often moderate.
Development companies are the primary users of the approach.
Their embedded software developers are especially interested in ensuring that the application code interacts correctly with the \ac{hal}.
The goal here is to prioritize interface requirements in order to single out the relevant ones.
Developers can then concentrate on the relevant interface requirements to write more reliable application code for embedded systems.
\Cref{fig:approach-process} illustrates the approach embodied as a process chart.
The approach comprises two parts.
Initially, we extract a candidate requirement from a given issue report and subsequently, we formally prove the relevance of the candidate requirement.
A detailed description of each part is provided below to explain the steps involved.

\subsubsection{Extract candidate from issue report.}
In the first part of the process, a developer extracts a candidate requirement from a given issue report on a system failure observed in a faulty embedded system using a \ac{hal} (see the blue box in \Cref{fig:approach-process}).
The extracted requirement serves as a candidate that may be responsible for the system failure, \ie, the requirement's violation captures the failure and the requirement's satisfaction witnesses the absence of the failure.

\paragraph{Inspect issue report.}
\label{sec:process-activity-analyze-issue-report}
In Activity~\num{1} (Manual inspection), the developer carefully analyzes the cause of the system failure from the issue report in the faulty system to create an error report.
Such a report should identify and describe the error, its underlying cause, and include details about the current configuration and state at failure.
If the issue report already contains these details, we can skip the activity and continue with Activity~\num{2}.
Finally, the activity produces a comprehensive error report.

\paragraph{Extract candidate requirement.}
\label{sec:process-activity-extract-requirement}
In Activity~\num{2} (Manual inspection), the developer extracts a candidate requirement based on the insights provided in the error report, especially focusing on the underlying cause that may indicate improper \ac{hal} usage.

To prevent improper \ac{hal} usage, the \ac{hal} functions must follow specific sequences to operate the hardware correctly.
These sequences can be prescribed by dependencies between \ac{hal} functions where certain functions depend on the successful execution of others.
To model this relationship, we use a binary relation representing a dependency between two \ac{hal} functions.
For a pair of \ac{hal} functions $f_1$ and $f_2$, the dependency $f_1() \opdep f_2()$ (\doublequote{$f_2$ depends on $f_1$}) expresses that a call of $f_1$ must precede a call of $f_2$ in the application code of an embedded system.
We call such a pair of \ac{hal} functions a \emph{temporal dependency}.

The developer carefully investigates the cause from the report to identify a dependency between \ac{hal} functions that may be related to the error.
Hereby, the developer consults the \ac{hal} documentation and if possible the \ac{hal} implementation to determine if a dependency exists.
Based on such a dependency, the developer extracts a temporal dependency as a candidate.
Finally, the activity produces the candidate requirement~$R$ for the relevance check.

\subsubsection{Check relevance of extracted candidate.}
\label{sec:process-activity-check-relevance}
The relevance check is the second part of the process (see the green box in \Cref{fig:approach-process}) and requires as input the candidate requirement~$R$.
The check involves two decisions (depicted in diamond shape in \Cref{fig:approach-process}) to witness the relevance of~$R$ according to \Cref{def:relevant-requirement}.
For the two decisions, we leverage a formal method, specifically software model checking (see, \eg, \cite{Heizmann2013}), to formally prove the relevance of~$R$.
For clarity, we explain below the two decisions of the relevance check.

\paragraph{First decision of relevance check.}
In the first step of the relevance check, we apply software model checking to the original application code of the faulty system, where we expect to detect a violation of~$R$.
If the decision returns the result \singlequote{yes}, a violation is found.
The violation indicates that~$R$ remains a candidate for relevance.
Contrarily, if no violation is detected, it implies that~$R$ is not responsible for the system failure from the issue report.

\let\oldintextsep\intextsep%
\setlength\intextsep{-\intextsep}%
\begin{wrapfigure}[25]{L}{6cm}
  \centering
  \input{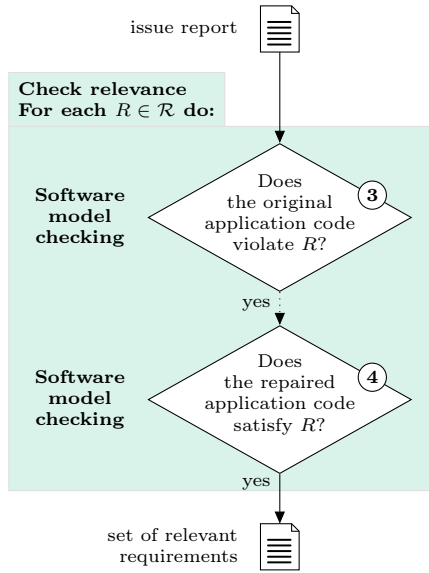}
  \caption{Alternative approach.
           The finite set of candidate requirements~$\mathcal{R}$ is fixed for a given \ac{hal} interface.
           The application code comes again in two versions: the original version exhibits the system failure, the repaired version does not.}
  \label{fig:approach-process-industrial}
\end{wrapfigure}
\setlength\intextsep{\oldintextsep}%

\paragraph{Second decision of relevance check.}
In the second step of the relevance check, we again apply software model checking, this time to the application code of the system in the repaired version.
Hereby, we aim to confirm that~$R$ is no longer violated.
Consequently, we expect that the application code in the repaired version satisfies~$R$.
If the decision returns the result \singlequote{yes}, the relevance check witnesses that~$R$ is indeed a relevant requirement.
Otherwise, if the decision returns the result \singlequote{no}, $R$ may be bogus.

\subsection{Check a set of candidate requirements for relevance}
\label{sec:approach-process-alternative}
We introduce an alternative approach for obtaining relevant interface requirements, embodied as a process chart in \Cref{fig:approach-process-industrial}.
This alternative approach of the relevance check is well-suited for use in industrial settings where companies collect issues in internal issue trackers.
Most of these issues have already been resolved, \ie, the corresponding faulty systems have been repaired and the reported system failures no longer occur.
But it remains unclear which requirements are relevant to prevent the system failures.

Here, the alternative approach can be applied retrospectively to single out relevant requirements.
Unlike the overall approach, this alternative does not require the manual extraction of a candidate requirement from a given issue report.
Instead, the alternative lifts the relevance check for a single candidate requirement (as shown in \Cref{fig:approach-process}) to a finite set of candidate requirements~$\mathcal{R}$.
The set~$\mathcal{R}$ is fixed for a given \ac{hal} interface and can be inferred from the \ac{hal} documentation and, if possible, from the \ac{hal} implementation.

Our formal notion of relevance allows us to leverage software model checking in order to simultaneously prove the relevance of each requirement~$R \in \mathcal{R}$, as explained in \Cref{sec:process-activity-check-relevance}.
If $R$ is proven relevant (\ie, both decisions return \singlequote{yes}) we add it to a finite set of relevant requirements.
Ideally, this set contains exactly one relevant requirement to emphasize the significance of relevance.
Multiple relevant requirements indicate that a system failure is masked by other failures.

  \section{Case study}
\label{sec:relevant-requirements}
In this section, we present a case study where we apply the approach presented in \Cref{sec:approach-process} to three issue reports on system failures.
Our goal is to evaluate the approach.
Specifically, we aim to address the following research questions:
\begin{enumerate}[%
    leftmargin=*,
    rightmargin=\parindent,
    labelindent=\parindent,
    labelsep=0.25em,
    label={\textbf{RQ\arabic*}:},
    itemsep=0.5em%
  ]
  \item Is the approach to extract a relevant \ac{hal} interface requirement from an issue report on a system failure feasible in principle?
  \item For each system failure, is there only one single requirement that is relevant for the system failure?
\end{enumerate}
The motivation behind RQ2 is our impression that the significance of relevance as a means to single out requirements is weakened in the case where two requirements are relevant for the same system failure.
To be able to give at least a preliminary answer to RQ2 we narrow down the question by restricting the set of requirements to the set of temporal dependencies that can be inferred for a given \ac{hal}.
This gives us a finite number of requirements and allows us to check each of them on the two versions of each system.

\subsection{System failures in embedded systems}
\label{sec:system-failures}
The system failures stem from issues that have recently received some attention in newsgroups, as discussed among several places including \cite{RaspberryPiForum2016}, \cite{RaspberryPiForum2019}, and~\cite{StackExchange2019}.
The issue reports concern embedded systems and provide structural system descriptions but they do not explicitly state interface requirements at all.

\subsubsection{Overview of systems.}
\begin{wrapfigure}{R}{0.6\linewidth}
  \centering
  \includegraphics[width=\linewidth]{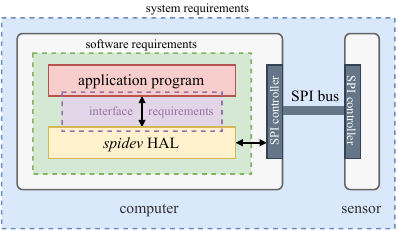}
  \caption{Architecture of the embedded system from the issue report in~\cite{RaspberryPiForum2019}.}
  \label{fig:embedded-system-structure}
\end{wrapfigure}

The embedded systems share a common structure, as illustrated in \Cref{fig:embedded-system-structure}, which specifically depicts the embedded system in~\cite{RaspberryPiForum2019}.
\let\oldintextsep\intextsep%
\setlength\intextsep{0pt}%
\setlength\intextsep{\oldintextsep}
While this schematic representation focuses on that particular example, it is important to note that the other two systems are similarly structured.

\paragraph{\acs{spi}~bus and \spidev \acs{hal}.}
The embedded systems use the \acf{spi}~\cite{Hill1990} as a communication bus to connect the system's hardware, in particular a computer or microcontroller, to a peripheral device.
The peripheral device is specifically an actuator like a display controller in~\cite{RaspberryPiForum2016}, a sensor in~\cite{RaspberryPiForum2019}, or an unspecified device in~\cite{StackExchange2019}.
The \ac{spi} bus is operated in a master/slave configuration where the integrated \ac{spi} controller of the computer or microcontroller serves as master to control the bus communication.
Conversely, the \ac{spi} controller of the peripheral device serves as slave.
The computer or microcontroller runs an application program that uses the \spidev \ac{hal}\footnote{The \spidev \ac{hal} with its userspace \ac{api} is part of the Linux kernel repository and available at \url{https://git.kernel.org/pub/scm/linux/kernel/git/torvalds/linux.git/tree/Documentation/spi/spidev?h=v5.3}} to operate the \ac{spi} bus via the master \ac{spi} controller (see \Cref{fig:embedded-system-structure}).
The \spidev \ac{hal} comprises several software layers to support \ac{spi} controllers from various manufacturers and uses \ac{posix} system calls to provide a \ac{hal} interface.
Proper operation of the \ac{spi} controller by the embedded application requires the use of \ac{hal} functions such as \halApiRoutineOpenS to initialize the \ac{spi} controller.
Configuration and data transfers are then performed through \ac{hal} functions like \halApiRoutineIoctlS, \halApiRoutineReadS, and \halApiRoutineWriteS.
The \halApiRoutineIoctlS function uses specific \ac{spi} constants%
\footnote{For reasons of readability, we have omitted the prefix \code{SPI\_IOC\_} for all \ac{spi} constants from the \spidev \acs{hal} and we have abbreviated the \ac{spi} constant \code{MESSAGE} with \code{MSG} (also see \Cref{tab:spi-hal-dependencies}).}
to set parameters such as mode, speed, and bit order for \ac{spi} data transfers.

\paragraph{Interface requirements.}
Interface requirements between the application program and the \spidev \ac{hal} (see dashed violate box in \Cref{fig:embedded-system-structure}) solely depend on the \ac{hal}.
These requirements have to be satisfied during the runtime of the system.
Otherwise runtime errors can occur that may lead to serious system failures.
Currently, such interface requirements do not exist for the \spidev \ac{hal}.
To address RQ1, we extract a candidate requirement for each of the three system failures according to our overall approach from \Cref{sec:approach-process-overall}.
To address RQ2, we systematically infer a set of requirements for the \spidev~\ac{hal} in order to apply our alternative approach from \Cref{sec:approach-process-alternative} retrospectively on the three systems.

\subsubsection{Replicated systems and corresponding failures.}
\label{sec:replicated-systems-failures}
While the reports offer some insights into the issues and the used embedded systems, they do not provide the complete application code or sufficient details of the hardware setup to clone each system.
We rebuild the embedded systems with the hardware to replicate system failures corresponding to those reported in the issue reports.
With the rebuild, we make sure that an error in each system lead to a system failure from the issue reports, \ie, that it is a real error (as opposed to a \emph{theoretical} error which cannot appear in a practical environment).
We create three applications that correspond to the applications from the issue reports: control an actuator (\cf~\cite{RaspberryPiForum2016}), read sensor measurements (\cf~\cite{RaspberryPiForum2019}), and test communication (\cf~\cite{StackExchange2019}).
In the following, we describe the three embedded systems and their observations that align with the ones from the issue reports.

\begin{embeddedsystem}{I/O~Expander}{es:io-expander}
{\textbf{System description}:}
The system \singlequote{I/O Expander} is designed to control a Microchip MCP23S17 input/output expander, which offers $16$~configurable input/output pins.
This system setup involves a master \ac{spi} controller on the computer and a slave \ac{spi} controller in the MCP23S17.
The application program uses the \spidev \ac{hal} to communicate with the MCP23S17, operating through a specified sequence of initialization and control commands.

\medskip{\textbf{Failure observation}:}
Upon executing the program, a system failure is observed in which pin~A0 of the MCP23S17 does not activate as expected.
This failure corresponds to the issue reported in \cite{RaspberryPiForum2016}, where the actuator (a display controller) fails to be controlled reliably.
\end{embeddedsystem}

\begin{embeddedsystem}{Accelerometer}{es:accelerometer}
{\textbf{System description}:}
The application \singlequote{Accelerometer} sends control commands to an Analog Devices ADXL345 accelerometer, which measures acceleration along three axes.
This system employs a similar master/slave configuration, where the computer acts as the \ac{spi} master and the ADXL345 acts as the \ac{spi} slave.
The application program retrieves sensor data by transmitting commands through the \spidev \ac{hal}, similar to how it might be executed in the original system.

\medskip{\textbf{Failure observation}:}
When running this application, we could observe a failure where no acceleration data can be received from the ADXL345.
This observation aligns with the system failure described in the issue report from \cite{RaspberryPiForum2019}, where a sensor fails to provide measurements.
\end{embeddedsystem}

\begin{embeddedsystem}{\spidev-Test}{es:spidev-test}
{\textbf{System description}:}
The application program \singlequote{\spidev-Test} is generally used by developers to test the functionality of newly developed \ac{spi} components.
In this case, we applied this application to test the communication with the ADXL345 accelerometer.
The system uses the same \ac{spi} master/slave configuration, with the computer configured to check the connectivity and data flow through the ADXL345.

\medskip{\textbf{Failure observation}:}
In executing this application, a failure occurs where the communication test fails to return a valid device \ac{id}.
This observes the issue reported in~\cite{StackExchange2019}, where the sanity check for communication with a peripheral device does not succeed.
\end{embeddedsystem}

\subsection{Extract and check candidate requirements from system failures}
We apply our overall approach from \Cref{sec:approach-process-overall} to the three issue reports in order to address RQ1.
We obtain the original application code from our replicated systems, where each system exhibits a failure.

\subsubsection{Inspect issue reports.}
The three issue reports describe system failures observed during operation and related to the \ac{spi} bus.
While the application code is syntactically correct and can be compiled and run, the issue reports indicate that the errors are not trivial.
Manual inspection of the system failures reveals that each application code properly initializes the \ac{spi} controller using the \ac{hal} function \halApiRoutineOpenS, but omits essential configuration calls, causing the controller to operate with default settings.
This misconfiguration leads to communication failures, as there is no universal agreement on a default configuration for \ac{spi} controllers.
In the \singlequote{I/O~Expander} code, the missing \halApiRoutineIoctlWSpeed call causes clock speed mismatches.
In the \singlequote{Accelerometer} code, the absent \halApiRoutineIoctlWModeExt call results in polarity and phase mismatches.
In the \singlequote{\spidev-Test} code, the missing \halApiRoutineIoctlWBitsPWord call causes word size mismatches in data transfers between the master and slave \ac{spi} controller.

\subsubsection{Extract candidate and check relevance.}
Following our overall approach, we extract a candidate requirement for each issue from the insights of the inspection.
Specifically, we extracted the temporal dependencies~$\varThad_{26}$ (for \singlequote{I/O~Expander}), $\varThad_{17}$ (for \singlequote{Accelerometer}), and $\varThad_{23}$ (for \singlequote{\spidev-Test}) as candidates (shown in detail in \Cref{tab:spi-hal-dependencies}).
We conduct the relevance check on the three candidates where we apply software model checking carried out by the verification tool \ultimateAutomizer~\cite{Heizmann2018} in version~0.3.0.\footnote{\ultimateAutomizer is part of the open-source program analysis framework \textsc{Ultimate} and available at \url{https://www.ultimate-pa.org}.}

\paragraph{Validity of temporal dependencies.}
\label{sec:checking-relevance:validity-properties}
Software model checking of the relevance check verifies the validity of extracted requirements within a C program that uses \spidev functions from the \ac{hal}.
We encode this validity through assertions in an annotated version of the program.
%

\setlength{\fboxsep}{1pt}
\begin{lstlisting}[
  mathescape,
  float,
  caption={Encoding the validity of a temporal dependency in an application program through the validity of an assertion in an annotation of the application program.
           Given the temporal dependency $\varThadDep{\varThad_{1}}{\halApiRoutineOpen}{\halApiRoutineRead}$ from Table~\ref{tab:spi-hal-dependencies}, the annotation consists in adding ghost code (in green) to the \spidev functions \halApiRoutineOpenS and \halApiRoutineReadS.
           The code snippets here refer to the original implementation of the \spidev functions from the \ac{hal}.
           The annotation does not modify the application program in other than the shown places.},
  label={lst:spi-hal-annotation}
]
<@\colorbox{green!20}{/*@\ \textbf{ghost} \textbf{int} state\_d1 = 0; */}@>

int open(const char *path, int oflag, ...)
{
    int ret = $\dots$;

    <@\colorbox{green!20}{/*@\ \textbf{ghost} state\_d1 = 1; */}@>
    return ret;
}

ssize_t read(int fd, void *buf, size_t nbyte)
{
    <@\colorbox{green!20}{/*@\ \textbf{assert} (state\_d1 == 1); */}@>

    return $\dots$;
}
\end{lstlisting}

The annotation, shown schematically in \Cref{lst:spi-hal-annotation}, adds auxiliary statements of two kinds: update statements that only modify auxiliary variables, and assert statements.
For each temporal dependency of the form~$\varThadDep{\varThad}{f_1()}{f_2()}$, we insert an update statement in the code for the implementation of function~$f_1$ and an assert statement in the function~$f_2$, in addition to the declaration of an auxiliary variable for the temporal dependency~$\varThad$.
In each execution, the value of the auxiliary variable (the \emph{ghost state}) flags whether the call of the function $f_1$ has (or has not yet) taken place.
We use the \ac{acsl}, an annotation language for~C, supported by a variety of verification tools~\cite{Baudin2024}.
The encoding of the validity is independent of the particular application code and solely depends on the \ac{hal}.

\paragraph{Experimental results.}
In the first step of the relevance check, we apply \ultimateAutomizer on each of the three application programs to verify the validity of the corresponding candidate temporal dependency.
The verification results reveal that one assert statement is violated in each faulty application program.
We inspect each violation (using the failure path from the verification result) to witness that the violated assert statements are indeed part of the program annotation.
Based on the program annotation, we do a reverse-lookup to locate the temporal dependencies, to which these violated assert statements belong.
The verification results witness that all three candidate temporal dependencies are violated in each corresponding application program.

In the second step of the relevance check, we again apply \ultimateAutomizer, this time to each of the application programs in the repaired version.
We obtain the repaired version by manual repair of the error in each faulty application program.
The failure path from each verification result can guide us to localize the errors, so that the repaired application programs no longer cause the observed system failures.
From the check, we obtain the verification results, which witness that all assertions in each of the three application programs are satisfied.
Thus, all three candidate temporal dependencies are satisfied and indeed relevant.
Our overall approach to obtain a relevant interface requirement succeeds in the three cases and is feasible in principle (\textbf{RQ1}).

\subsection{Check a set of \spidev \ac{hal} interface requirements for relevance}
\let\oldintextsep\intextsep%
\setlength\intextsep{-\intextsep}%
\begin{wraptable}{R}{5.3cm}
  \centering
  \caption{Candidate temporal dependencies inferred from the \spidev \acs{hal}.}
  \label{tab:spi-hal-dependencies}
  \medskip
  \scriptsize

\setlength{\tabcolsep}{1pt}
\newcommand{\varThadDepRowS}[3]{#1 & $:$ & #2 \hspace*{1.4em} & $\opdep$ & \multicolumn{3}{l}{#3} \\}
\newcommand{\varThadDepRowL}[3]{#1 & $:$ & \multicolumn{3}{l}{#2} & $\opdep$ & #3 \\}
\begin{tabular}{lclclcl}
  \toprule
  \varThadDepRowS{$\varThad_{1}$} {$\halApiRoutineOpen$}{$\halApiRoutineRead$}
  \varThadDepRowS{$\varThad_{2}$} {$\halApiRoutineOpen$}{$\halApiRoutineWrite$}
  \varThadDepRowS{$\varThad_{3}$} {$\halApiRoutineOpen$}{$\halApiRoutineIoctlMsg$}
  \varThadDepRowS{$\varThad_{4}$} {$\halApiRoutineOpen$}{$\halApiRoutineClose$}
  \varThadDepRowS{$\varThad_{5}$} {$\halApiRoutineOpen$}{$\halApiRoutineIoctlRMode$}
  \varThadDepRowS{$\varThad_{6}$} {$\halApiRoutineOpen$}{$\halApiRoutineIoctlWMode$}
  \varThadDepRowS{$\varThad_{7}$} {$\halApiRoutineOpen$}{$\halApiRoutineIoctlRModeExt$}
  \varThadDepRowS{$\varThad_{8}$} {$\halApiRoutineOpen$}{$\halApiRoutineIoctlWModeExt$}
  \varThadDepRowS{$\varThad_{9}$} {$\halApiRoutineOpen$}{$\halApiRoutineIoctlRLSBF$}
  \varThadDepRowS{$\varThad_{10}$}{$\halApiRoutineOpen$}{$\halApiRoutineIoctlWLSBF$}
  \varThadDepRowS{$\varThad_{11}$}{$\halApiRoutineOpen$}{$\halApiRoutineIoctlRBitsPWord$}
  \varThadDepRowS{$\varThad_{12}$}{$\halApiRoutineOpen$}{$\halApiRoutineIoctlWBitsPWord$}
  \varThadDepRowS{$\varThad_{13}$}{$\halApiRoutineOpen$}{$\halApiRoutineIoctlRSpeed$}
  \varThadDepRowS{$\varThad_{14}$}{$\halApiRoutineOpen$}{$\halApiRoutineIoctlWSpeed$}
  \midrule
  \varThadDepRowL{$\varThad_{15}$}{$\halApiRoutineIoctlWModeExt$}  {$\halApiRoutineRead$}
  \varThadDepRowL{$\varThad_{16}$}{$\halApiRoutineIoctlWModeExt$}  {$\halApiRoutineWrite$}
  \varThadDepRowL{$\varThad_{17}$}{$\halApiRoutineIoctlWModeExt$}  {$\halApiRoutineIoctlMsg$}
  \varThadDepRowL{$\varThad_{18}$}{$\halApiRoutineIoctlWLSBF$}     {$\halApiRoutineRead$}
  \varThadDepRowL{$\varThad_{19}$}{$\halApiRoutineIoctlWLSBF$}     {$\halApiRoutineWrite$}
  \varThadDepRowL{$\varThad_{20}$}{$\halApiRoutineIoctlWLSBF$}     {$\halApiRoutineIoctlMsg$}
  \varThadDepRowL{$\varThad_{21}$}{$\halApiRoutineIoctlWBitsPWord$}{$\halApiRoutineRead$}
  \varThadDepRowL{$\varThad_{22}$}{$\halApiRoutineIoctlWBitsPWord$}{$\halApiRoutineWrite$}
  \varThadDepRowL{$\varThad_{23}$}{$\halApiRoutineIoctlWBitsPWord$}{$\halApiRoutineIoctlMsg$}
  \varThadDepRowL{$\varThad_{24}$}{$\halApiRoutineIoctlWSpeed$}    {$\halApiRoutineRead$}
  \varThadDepRowL{$\varThad_{25}$}{$\halApiRoutineIoctlWSpeed$}    {$\halApiRoutineWrite$}
  \varThadDepRowL{$\varThad_{26}$}{$\halApiRoutineIoctlWSpeed$}    {$\halApiRoutineIoctlMsg$}
  \bottomrule
\end{tabular}

\end{wraptable}
\setlength\intextsep{\oldintextsep}%
We apply our alternative approach from \Cref{sec:approach-process-alternative} retrospectively to the three issue reports in order to address RQ2.
We obtain the application code in the original, faulty version from the replicated systems, and in the repaired, correct version from our experiments addressing RQ1.

\subsubsection{Infer a set of candidates.}
The approach requires a set of interface requirements, for which we infer temporal dependencies from the \spidev \ac{hal}.
We use a systematic method to infer the temporal dependencies where we investigate all pairs of \ac{hal} functions.
For each pair, we consult the documentation of the \ac{hal} and, if possible, the \ac{hal} implementation to determine if a dependency exists.
Confirmed dependencies are added as temporal dependencies to the set.
Finally, we obtain the set of temporal dependencies shown in \Cref{tab:spi-hal-dependencies}.

\subsubsection{Check relevance of candidates.}
For each of the three issue reports, we apply the relevance check to the set of temporal dependencies from \Cref{tab:spi-hal-dependencies}.
We again utilize \ultimateAutomizer, this time to simultaneously verify the validity of all temporal dependencies from the set.
This requires that we encode the validity of all temporal dependencies as annotation in both versions of each application program.

\paragraph{Experimental results.}
\begin{table}
  \centering
  \caption{Experimental results of the applied relevance check to the inferred temporal dependencies for the \spidev \acs{hal} from \Cref{tab:spi-hal-dependencies}.
           The set listed in under \singlequote{Result} is exactly the set of interface requirements that the check determines to be violated.
           That is, the complement contains those that the check determines to be correct.}%
  \label{tab:case-study-results}
  \scriptsize

\newcommand{\hdo}[1]{\textbf{#1}}
\newcommand{\hdt}[1]{\scriptsize\textbf{#1}}
\newcommand{\loc}[1]{\numunit{#1}}
\newcommand{\pfault}{\scriptsize faulty}
\newcommand{\pcorrt}{\scriptsize repaired}
\newcommand{\uloc}{\tiny [\ac{loc}]}
\renewcommand{\us}{\tiny [s]}
\newcommand{\umb}{\tiny [MB]}
\setlength{\tabcolsep}{14pt}
\begin{tabular}{llrcrr}
  \toprule
  \belowrulesepcolor{bgray}
  \rowcolor{bgray} \multicolumn{3}{c|}{\hdo{Application program}}                                                       & \hdo{Result}          & \multicolumn{2}{|c}{\hdo{Verification}} \\
  \rowcolor{bgray} \multicolumn{1}{c}{\hdt{name}} & \multicolumn{1}{c}{\hdt{version}} & \multicolumn{1}{c|}{\hdt{size}} & \hdt{incorrect}       & \multicolumn{1}{|c}{\hdt{time}} & \multicolumn{1}{c}{\hdt{mem.}} \\[-0.2em]
  \rowcolor{bgray}                                &                                   & \multicolumn{1}{c|}{\uloc}      &                       & \multicolumn{1}{|c}{\us}        & \multicolumn{1}{c}{\umb} \\[-0.2em]
  \aboverulesepcolor{bgray}
  \midrule
  \multirow{2}{*}{I/O Expander}                   & \pfault                           & \multicolumn{1}{|r|}{\loc{172}} & $\set{\varThad_{26}}$ & \multicolumn{1}{|r|}{\numunit{6.24}}   & \numunit{264} \\
                                                  & \pcorrt                           & \multicolumn{1}{|r|}{\loc{180}} & $\emptyset$           & \multicolumn{1}{|r|}{\numunit{8.10}}   & \numunit{263} \\
  \midrule
  \multirow{2}{*}{Accelerometer}                  & \pfault                           & \multicolumn{1}{|r|}{\loc{284}} & $\set{\varThad_{17}}$ & \multicolumn{1}{|r|}{\numunit{10.43}}  & \numunit{294}  \\
                                                  & \pcorrt                           & \multicolumn{1}{|r|}{\loc{292}} & $\emptyset$           & \multicolumn{1}{|r|}{\numunit{12.36}}  & \numunit{345}  \\
  \midrule
  \multirow{2}{*}{\spidev-Test}                   & \pfault                           & \multicolumn{1}{|r|}{\loc{764}} & $\set{\varThad_{23}}$ & \multicolumn{1}{|r|}{\numunit{990.35}} & \numunit{1129} \\
                                                  & \pcorrt                           & \multicolumn{1}{|r|}{\loc{768}} & $\emptyset$           & \multicolumn{1}{|r|}{\numunit{925.58}} & \numunit{1335} \\
  \bottomrule
\end{tabular}

\end{table}
\Cref{tab:case-study-results} shows the experimental results of the relevance check.
Three out of the total~$26$ candidate temporal dependencies (namely~$\varThad_{17}$, $\varThad_{23}$, and~$\varThad_{26}$) are formally proven relevant.
These three requirements are responsible for the system failures from the three issue reports.
For each system failure, we obtain exactly one relevant interface requirement (\textbf{RQ2}).
This finding emphasizes the significance of relevance in our case study.

\subsection{Discussion}
Our results demonstrate that we obtain the same relevant interface requirements using our overall approach as well as the alternative relevance check.
This consistency provides a strong indication that our approach is feasible in principle.
In our case study, we did not encounter the scenario where one system failure masks other ones, which indicates that several requirements might be relevant.
Masking failures could occur if a developer forgets to insert not only the \ac{hal} function calls for the configuration of the master \ac{spi} controller but also the call to initialize the \ac{spi} controller, leading to several observable system failures that may mask one another.
While this case could potentially arise, our alternative relevance check from \Cref{sec:approach-process-alternative} can obtain several relevant requirements even in the presence of masking failures.

The experimental results from \Cref{tab:case-study-results} also indicate that the verification of the validity of temporal dependencies itself is not a challenge in this work (\cf last two columns of \Cref{tab:case-study-results}).
Even the more complex application program \singlequote{\spidev-Test} with $766$~lines of code (including the annotation) in its repaired version can be verified within \numunit{17}{min} where the memory consumption never exceeds \numunit{1.4}{GB}.\footnote{All verification runs with \ultimateAutomizer were carried out and measured on a regular desktop computer with quad-core \ac{cpu} at \numunit{2.6}{GHz}, \numunit{8}{GB} of memory, running Arch~Linux with Linux kernel~6.12.47.}
For this reason, it does not seem useful to evaluate the efficiency of other software model checking tools or formal verification methods.

Several factors could affect the validity of our findings from the experimental evaluation.
First, the representativeness of the observed system failures is crucial, as it influences the relevance of the interface requirements.
The lack of case studies on system failures and the often imprecise descriptions limit our understanding of potential issues.
Due to the unavailability of the original systems, we replicated comparable embedded systems to reproduce similar system failures.
Consequently, the findings from these replicated systems cannot be directly applied to the original systems; instead, they provide evidence that the independently identified interface requirements may be relevant for failures resembling those in the issue reports.
In industrial practice, the availability of the actual systems would reduce the importance of system replication to obtain application code.
More research is needed to confirm the generality of our results.


\section{Related work}
\label{sec:related-work}
In this section, we cover work related to the topics in this paper.
We focus on areas such as requirement prioritization, interface requirements, and methods for deriving requirements.

\paragraph{Requirement prioritization.}
Recent surveys~\cite{Bukhsh2020,Hujainah2018,Pacheco2018,Yaseen2025}, provide an overview of requirements prioritization techniques, with guidelines to choose a suitable prioritization technique~\cite{CastilloValdivieso2024}.
While most techniques rely on manual effort, our formal notion of relevance allows us to leverage a formal method to automatically check relevance.
A related review study~\cite{Riegel2015} focuses on the most important aspect of single out requirements, the prioritization criteria.
However, most criteria focus on importance, cost-benefit, and stakeholder satisfaction and leave system reliability and safety out of consideration.
In addition, many criteria are ambiguous and not clearly understandable leading to subjective and inconsistent prioritization decisions.
In contrast, this work introduces a formal relevance criterion specifically designed to address system reliability and safety.
Its formal nature ensures unambiguity and clarity, which helps to single out important interface requirements that are essential for preventing system failures.

\paragraph{Temporal dependencies.}
The class of temporal dependencies which we used to define the set of \num{26} candidate requirements corresponds to the precedence requirement pattern in the hierarchy in~\cite{Dwyer1999}.
The class is restrictive enough to enable a systematic selection of candidate requirements; yet, it is expressive enough to include relevant requirements, as shown in our case study.

\paragraph{Deriving requirements.}
Previous work~\cite{Beyer2007}, building upon the work in~\cite{Alur2005} and~\cite{Henzinger2005}, proposes different algorithms to automatically derive interface requirements from an existing library implementation for its safe and permissive use.
These approaches require the presence of explicit safety requirements (\eg, invariants) within the library implementation, which serve as a basis for the automatic inference of a possibly comprehensive set of interface requirements.
In contrast, our approach does not depend on safety requirements defined a priori.
Instead, the primary objective of our approach is to derive a concise set of \ac{hal} interface requirements by prioritizing candidate requirements to single out the relevant ones that are essential for preventing system failures.
A similar line of research is the specification mining based on statistical or stochastic methods where interface requirements are automatically derived from a set of error-free application programs that properly interact with the interface, \eg, see~\cite{Amann2016,Amann2019}.
The mined specifications are utilized to detect so-called \ac{api}~(mis)uses.
The existing approaches to automatically mine interface requirements may be contrasted with our approach to derive relevant \ac{hal} interface requirements: the former relies on programs that properly use a given interface, whereas the latter relies in the existence of programs that do not.
Exploring how one can exploit this apparent complementariness remains an open area for future research.


\section{Conclusion}
\label{sec:conclusion}
In this work, we introduced a formal notion of a relevance criterion for requirements prioritization based on system failures.
The criterion allows us to single out those requirements that are essential for preventing system failures.
Its formal notion ensures a clear understanding making the prioritization consistent, unambiguous, and independent of any subjective assessment.

We presented an approach to prioritize requirements using the example of \ac{hal} interface requirements by formally proving their relevance within a defined process.
\ac{hal} interface requirements define the correct use of a \ac{hal} interface in application programs for embedded systems.
Our approach for obtaining relevant \ac{hal} interface requirements starts by analyzing issue reports that mention observable system failures.
From these reports, we infer candidate requirements for the correct use of the \ac{hal} interface.
We then formally prove the relevance of each candidate requirement by leveraging software model checking.
This relevance check is an essential process step, as it witnesses that an interface requirement is indisputably responsible for an observed system failure.

We evaluated our approach in a case study on three examples of issue reports related to embedded application programs that use the \ac{spi} bus via the \spidev \ac{hal}.
The experimental results demonstrate the practical feasibility of our approach.
We were able to obtain a relevant interface requirement for each mentioned system failure from the three issue reports.
Each relevant interface requirements is essential in preventing an observable system failure.
Embedded software developers should concentrate on such relevant requirements to prevent specific kinds of system failures, such as those mentioned in the issue reports.


\begin{credits}
\subsubsection{\ackname}
Part of this study was funded by the Deutsche Forschungsgemeinschaft (DFG, German Research Foundation) -- \censor{503812980}.
\end{credits}

  \bibliographystyle{splncs04}
  \bibliography{literature/references}

@Article{Hujainah2018,
  author    = {Hujainah, Fadhl and Bakar, Rohani Binti Abu and Abdulgabber, Mansoor Abdullateef and Zamli, Kamal Z.},
  journal   = {IEEE Access},
  title     = {Software Requirements Prioritisation: A Systematic Literature Review on Significance, Stakeholders, Techniques and Challenges},
  year      = {2018},
  pages     = {71497--71523},
  volume    = {6},
  doi       = {10.1109/access.2018.2881755},
  publisher = {IEEE},
}

@Article{Bukhsh2020,
  author    = {Bukhsh, Faiza Allah and Bukhsh, Zaharah Allah and Daneva, Maya},
  journal   = {Computer Standards \& Interfaces},
  title     = {A systematic literature review on requirement prioritization techniques and their empirical evaluation},
  year      = {2020},
  month     = mar,
  pages     = {103389},
  volume    = {69},
  doi       = {10.1016/j.csi.2019.103389},
  publisher = {Elsevier},
}

@Article{Yaseen2025,
  author    = {Yaseen, Muhammad and Mehmood, Waqar and Hameed, Fazal and Nauman, Muhammad Asif},
  journal   = {Journal of Software: Evolution and Process},
  title     = {Scalability and Limitations of Existing Software Requirements Prioritization Techniques: A Systematic Literature Review},
  year      = {2025},
  month     = aug,
  number    = {8},
  volume    = {37},
  doi       = {10.1002/smr.70039},
  publisher = {Wiley},
}

@InProceedings{Heizmann2013,
  author    = {Matthias Heizmann and Jochen Hoenicke and Andreas Podelski},
  booktitle = {Computer Aided Verification},
  title     = {Software Model Checking for People Who Love Automata},
  year      = {2013},
  pages     = {36--52},
  publisher = {Springer},
  series    = {CAV},
  doi       = {10.1007/978-3-642-39799-8_2},
}

@Misc{RaspberryPiForum2016,
  howpublished = {Raspberry~Pi Forum},
  title        = {\acs{spi} connection between {RPi} and {ATmega32}},
  year         = {2016},
  key          = {Raspberry~Pi Forum},
  url          = {https://forums.raspberrypi.com/viewtopic.php?p=881605},
  urldate      = {2024-05-24},
}

@Misc{RaspberryPiForum2019,
  howpublished = {Raspberry~Pi Forum},
  title        = {Reading values from {MCP3002} not working [\dots]},
  year         = {2019},
  key          = {Raspberry~Pi Forum},
  url          = {https://forums.raspberrypi.com/viewtopic.php?t=230569},
  urldate      = {2024-05-24},
}

@Misc{StackExchange2019,
  howpublished = {Stack Exchange},
  title        = {\spidev sanity check not working},
  year         = {2019},
  key          = {Stack Exchange},
  url          = {https://raspberrypi.stackexchange.com/questions/102286/spidev-sanity-check-not-working},
  urldate      = {2024-05-24},
}

@Misc{Hill1990,
  author       = {Susan C. Hill and Joseph Jelemensky and Mark R. Heene and Stanley E. Groves and Daniel N. DeBrito},
  howpublished = {Patent~US-4958277},
  month        = sep,
  title        = {Queued Serial Peripheral Interface For Use In A Data Processing System},
  year         = {1990},
  address      = {United States Patent},
  day          = {18},
  dayfiled     = {21},
  filing_num   = {342651},
  ipc_class    = {G06F 3/00},
  monthfiled   = {apr},
  number       = {US-4958277},
  type         = {patent},
  url          = {https://image-ppubs.uspto.gov/dirsearch-public/print/downloadPdf/4958277},
  urldate      = {2024-05-23},
  us_class     = {364/200},
  yearfiled    = {1989},
}

@InProceedings{Heizmann2018,
  author    = {Matthias Heizmann and Yu-Fang Chen and Daniel Dietsch and Marius Greitschus and Jochen Hoenicke and Yong Li and Alexander Nutz and Betim Musa and Christian Schilling and Tanja Schindler and Andreas Podelski},
  booktitle = {Tools and Algorithms for the Construction and Analysis of Systems},
  title     = {{\ultimateAutomizer} and the Search for Perfect Interpolants (Competition Contribution)},
  year      = {2018},
  editor    = {Dirk Beyer and Marieke Huisman},
  pages     = {447--451},
  publisher = {Springer},
  series    = {TACAS},
  doi       = {10.1007/978-3-319-89963-3_30},
}

@Manual{Baudin2024,
  title        = {ACSL: ANSI/ISO C Specification Language},
  author       = {Baudin, Patrick and Cuoq, Pascal and Filliâtre, Jean-Christophe and Marché, Claude and Monate, Benjamin and Moy, Yannick and Prevosto, Virgile},
  note         = {Version 1.20},
  organization = {Frama-C},
  year         = {2024},
  url          = {https://frama-c.com/download/acsl-1.20.pdf},
  urldate      = {2024-04-25},
}

@Article{Pacheco2018,
  author    = {Pacheco, Carla and García, Ivan and Reyes, Miryam},
  journal   = {IET Software},
  title     = {Requirements elicitation techniques: a systematic literature review based on the maturity of the techniques},
  year      = {2018},
  month     = aug,
  number    = {4},
  pages     = {365--378},
  volume    = {12},
  doi       = {10.1049/iet-sen.2017.0144},
  publisher = {Institution of Engineering and Technology},
}

@Article{CastilloValdivieso2024,
  author    = {Castillo-Valdivieso, Pedro A. and Alhenawi, Esraa and Awawdeh, Shatha and Abu Khurma, Ruba and García-Arenas, Maribel and Hudaib, Amjad},
  journal   = {Journal of Computer Science \& Technology},
  title     = {Choosing a Suitable Requirement Prioritization Method: A Survey},
  year      = {2024},
  month     = apr,
  number    = {1},
  pages     = {e04},
  volume    = {24},
  doi       = {10.24215/16666038.24.e04},
  publisher = {Universidad Nacional de La Plata},
}

@InProceedings{Riegel2015,
  author    = {Riegel, Norman and Doerr, Joerg},
  booktitle = {Requirements Engineering: Foundation for Software Quality},
  title     = {A Systematic Literature Review of Requirements Prioritization Criteria},
  year      = {2015},
  pages     = {300--317},
  publisher = {Springer},
  series    = {REFSQ},
  doi       = {10.1007/978-3-319-16101-3_22},
}

@InProceedings{Dwyer1999,
  author    = {Dwyer, Matthew B. and Avrunin, George S. and Corbett, James C.},
  booktitle = {Proceedings of the International Conference on Software Engineering},
  title     = {Patterns in Property Specifications for Finite-state Verification},
  year      = {1999},
  pages     = {411--420},
  publisher = {ACM},
  series    = {ICSE},
  doi       = {10.1145/302405.302672},
}

@InProceedings{Beyer2007,
  author    = {Beyer, Dirk and Henzinger, Thomas A. and Singh, Vasu},
  booktitle = {Computer Aided Verification},
  title     = {Algorithms for Interface Synthesis},
  year      = {2007},
  editor    = {Damm, Werner and Hermanns, Holger},
  pages     = {4--19},
  publisher = {Springer},
  series    = {CAV},
  doi       = {10.1007/978-3-540-73368-3_4},
}

@Article{Alur2005,
  author    = {Alur, Rajeev and Černý, Pavol and Madhusudan, P. and Nam, Wonhong},
  journal   = {SIGPLAN Notices},
  title     = {Synthesis of interface specifications for {Java} classes},
  year      = {2005},
  number    = {1},
  pages     = {98--109},
  volume    = {40},
  doi       = {10.1145/1047659.1040314},
  publisher = {ACM},
}

@Article{Henzinger2005,
  author    = {Thomas A. Henzinger and Ranjit Jhala and Rupak Majumdar},
  journal   = {Software Engineering Notes},
  title     = {Permissive interfaces},
  year      = {2005},
  number    = {5},
  pages     = {31--40},
  volume    = {30},
  doi       = {10.1145/1095430.1081713},
  publisher = {ACM},
  series    = {FSE},
}

@InProceedings{Amann2016,
  author    = {Sven Amann and Sarah Nadi and Hoan A. Nguyen and Tien N. Nguyen and Mira Mezini},
  booktitle = {Proceedings of the International Conference on Mining Software Repositories},
  title     = {{MUBench}: a benchmark for {API}-misuse detectors},
  year      = {2016},
  publisher = {ACM},
  series    = {MSR},
  doi       = {10.1145/2901739.2903506},
}

@Article{Amann2019,
  author    = {Sven Amann and Hoan Anh Nguyen and Sarah Nadi and Tien N. Nguyen and Mira Mezini},
  journal   = {Transactions on Software Engineering},
  title     = {A Systematic Evaluation of Static {API}-Misuse Detectors},
  year      = {2019},
  number    = {12},
  pages     = {1170--1188},
  volume    = {45},
  doi       = {10.1109/tse.2018.2827384},
  publisher = {IEEE},
  series    = {TSE},
}

\end{document}